\documentclass[aps,prb,twocolumn,groupedaddress]{revtex4}
\usepackage{graphicx}
\usepackage{wasysym}
\usepackage{bm}
\usepackage{color}
\usepackage{ulem}

\begin{document}
\title{ Three-dimensional quantum Hall effect and magnetothermoelectric properties in Weyl semimetals}
\author{R. Ma$^{1,2}$}
\email{njrma@hotmail.com}
\author{D. N. Sheng$^{2}$}
\email{donna.sheng@csun.edu}
\author{L. Sheng$^{3,4}$}
\email{shengli@nju.edu.cn}

\address{$^1$ Jiangsu Key Laboratory for Optoelectronic Detection of Atmosphere and Ocean,
  Nanjing University of Information Science and Technology, Nanjing 210044, China\\
$^2$ Department of Physics and Astronomy, California State
University, Northridge, California 91330, USA\\
$^3$ National Laboratory of Solid State Microstructures and
Department of Physics, Nanjing University, Nanjing 210093, China\\
$^4$ Collaborative Innovation Center of Advanced Microstructures,
Nanjing 210093, China}

\begin{abstract}
We numerically study the three-dimensional (3D) quantum Hall effect (QHE)
and magnetothermoelectric transport of Weyl semimetals in the
presence of disorder. We obtain a bulk picture that the exotic 3D QHE
emerges in a finite range of Fermi energy around  the Weyl points determined
by the gap between the $n=-1$ and $n=1$ Landau levels (LLs).
The quantized Hall conductivity is attributable to  the chiral zeroth LLs
traversing the gap, and is robust against disorder scattering
for an intermediate number of layers in the direction
of the magnetic field. Moreover, we predict several interesting
characteristic features of the thermoelectric transport coefficients
in the 3D QHE regime, which can be probed experimentally.
This may open a new avenue for exploring Weyl physics in topological materials.
\end{abstract}

\maketitle

\section{Introduction}
\label{sec:intro}

Weyl semimetals have been attracting intense  interests
in recent years~\cite{Wan2011,Weng2015,Lv2015,Jiang2015,Chen2015,Nielsen,Huang2015,Xu2015,Lv15,Yang2015}.
The band touching points known as the Weyl points always appear in pairs with the opposite chirality~\cite{Nielsen},
and act like magnetic monopoles in momentum space with quantized Berry flux.
Another prominent feature of Weyl semimetals is the existence
of topologically protected surface states.
These surface states in momentum space form nonclosed Fermi arc,
connecting the Weyl points projected to the surface Brillouin zone.
Due to these unique features,
Weyl semimetals exhibit many exotic quantum transport properties, such as
chiral anomaly~\cite{Nielsen,yang2011,Vazifeh2013,Parameswaran2014},
the accompanying negative magnetoresistance~\cite{Son2013,Burkov2014,Lu2015},
and the planar Hall effect~\cite{Nandy2017,Burkov2017,Kumar2018,Yang2019}.
In particular, the three-dimensional (3D) quantum Hall effect (QHE)
is predicted to occur in Weyl semimetals, where the Fermi arcs
at opposite surfaces can form a complete Fermi loop and support the QHE
by a ``wormhole'' tunneling between the Weyl points~\cite{Wang2017,Lu2019,Sun2019}.
As it is  well known, 3D systems normally do not exhibit the QHE
owing to the continuum  spectrum from the
band dispersion along the direction of the magnetic field.
Therefore, such an intriguing transport signature of Weyl semimetals
has attracted more research to reveal the physics of
the 3D QHE~\cite{Zhang2017,Zhang2019,Lin2019,Uchida2017,Gao2018,Li2020}.
Experimentally, the 3D QHE was observed in Dirac semimetal ZrTe$_5$ crystals~\cite{Tang2019},
and a charge-density-wave mechanism of the 3D QHE is also proposed  to explain  experimental
observations~\cite{Qin2020}.
However, the interplay between the system sizes, the magnetic field strength
and the disorder scattering effect of the 3D QHE in Weyl semimetals have not been
understood.

Another exciting frontier is to explore  the thermoelectric transport of Weyl semimetals,
due to the possibility of record-high thermoelectric conversion efficiency in these
semimetal systems
~\cite{Caglieris2018,Watzman2018,Hirschberger2016,Chen2016,
Gooth2017,Sharma2016,Gorbar2017,Yang18,Jia2016}.
More recently, a nonsaturating thermopower and quantized thermoelectric
Hall conductivity has been proposed for Weyl semimetals~\cite{Fu1904,Skinner2018,Fu2019}.
Although there has been much work on the thermoelectric transport
properties, thermoelectric transport in the 3D QHE regime
and the effect of the disorder scattering have not been studied.
Such an investigation is highly desired.

In this paper, we report a numerical study of the QHE and magnetothermoelectric
transport of a 3D Weyl semimetal in the presence of disorder.
We demonstrate that the Hall conductivity $\sigma_{xy}$ exhibits
well-defined plateaus in units of $e^{2}/h$ for electron Fermi energy in the
finite gap $E_g$ ($E_g = 2\sqrt{2}\hbar \omega_c$ with $\omega_c$ as the cyclotron frequency)
between the $n = -1$ and $n = 1$ Landau levels (LLs)
around the Weyl points, as the zeroth LLs are discretized for an intermediate number of
layers in the direction of the magnetic field.
Our theory suggests a new version of QHE protected by bulk energy gap,
which may occur in the 3D Weyl semimetal, and is different from the 3D QHE
based on ``wormhole'' tunneling proposed in Refs.21-23.
We show how the system size, the magnetic field strength
and disorder influence the quantized Hall plateaus.
We further reveal that the transverse thermoelectric conductivity
$\alpha_{xy}$ develops a plateau for a wide range of temperatures,
which is quantized at an universal constant signaling the LL quantization.
The Nernst signal $S_{xy}$ shows a broad maximum at intermediate $T$ for strong magnetic fields,
which shifts to lower $T$ with decreasing magnetic field strength.
Our work provides a systematical understanding of the topological
3D QHE and magnetothermoelectric transport in Weyl semimetals.

This paper is organized as follows.
In Sec. II, the model Hamiltonian of the 3D Weyl semimetal is introduced.
In Sec. III, numerical results of the quantized Hall conductivity and thermoelectric transport
coefficients obtained by using exact diagonalization
are presented. The final section contains a summary.

\section{Model and Methods}

Let us start from a minimal two-band model of Weyl semimetals
on a 3D cubic lattice, whose Hamiltonian in momentum space
is given by~\cite{Armitage2018}
\begin{eqnarray}
H&=&t_x(\sin k_xa)\sigma_x+t_y(\sin k_ya)\sigma_y\nonumber\\
&+&(M_1-t_x\cos k_xa-t_y\cos k_ya-t_z\cos k_za)\sigma_{z},\label{TBHMomentum}
\end{eqnarray}
where $a$ is the lattice constant,
and $t_i$ ($i=x,y,z$) denotes the hopping integral along the $i$ axis.
$\mbox{\boldmath{$\sigma$}}=(\sigma_x,\sigma_y,\sigma_z)$
is the Pauli matrices for the pseudospin orbital degrees of freedom.
${\bf k}=(k_x,k_y,k_z)$ is the wave vector,
and $M_1$ is the effective Zeeman strength.
For $\vert(M_1-t_x-t_y)/t_z\vert<1$, as considered here,
a pair of Weyl points are located at ${\bf k_\pm}=(0,0,\pm k_0)$
with $\cos(k_0a) = (M_1-t_x-t_y)/t_z$. Expanding Eq.\ (\ref{TBHMomentum})
around the two Weyl points, one can obtain  the corresponding
low-energy effective Hamiltonian
$H=\hbar v_F^x\sigma_x q_x+\hbar v_F^y\sigma_y q_y\pm\hbar v_F^z\sigma_z q_z$,
where $\pm$ are for the two Weyl valleys, and ${\bf q} ={\bf k}-{\bf k}_{\pm}$ are
the relative wave vectors.  $v_{F}^{x}=t_xa/\hbar$, $v_{F}^{y}=t_ya/\hbar$, and $v_{F}^{z}=t_za\sin(k_0a)/\hbar$
are the Fermi velocities.

In real space, when a homogeneous magnetic field ${\bf B}=(0,0,B)$
is applied along the $z$ direction,
the tight-binding Hamiltonian on the cubic lattice corresponding to
Eq.\ (\ref{TBHMomentum}) can be written as
~\cite{Gao2018},
\begin{eqnarray}
H&=&\sum\limits_{\langle {nml}\rangle}C_{n+1,m,l}^{\dagger}T_{x}C_{n,m,l}
+e^{n(2\pi\phi i)}C_{n,m+1,l}^{\dagger}T_{y}C_{n,m,l}\nonumber\\
&+&C_{n,m,l+1}^{\dagger
}T_{z}C_{n,m,l}+\frac{M_1}{2}C_{n,m,l}^{\dagger
}\sigma_{z}C_{n,m,l}\nonumber\\
&+&w_{n,m,l}C_{n,m,l}^{\dagger}C_{n,m,l}+H.c.\ .
\end{eqnarray}
Here, the summation of $\langle {nml}\rangle$
runs over all lattice sites.
$C_{n,m,l}^{\dagger}=(C_{n,m,l,\uparrow}^{\dagger},C_{n,m,l,\downarrow}^{\dagger})$
is the two-component creation operators of electrons
on the lattice site with coordinates $(n,m,l)$ along the ${x,y,z}$ direction, respectively.
$T_x=-\frac{1}{2}t_x(i\sigma_x+\sigma_z)$,
$T_y=-\frac{1}{2}t_y(i\sigma_y+\sigma_z)$ and $T_z=-\frac{1}{2}t_z\sigma_z$
denote the $2\times2$ hopping matrices along the three directions, respectively.
$\phi$ stands for the magnetic flux
per square in units of flux quantum $\phi_0 = h/e$, namely, $\phi=Ba^2/\phi_0$. The  magnetic field strength $B$
is determined by $B= \phi\phi_{0}/a^2$. Since we focus on the low-energy or equivalently
long-wavelength properties of the model, the results are insensitive to the
lattice structure.
In order to obtain a realistic magnetic field $B$,
we may choose a relatively large lattice constant $a$.
For example, we choose $a=20\AA$, and then $\phi = 1/192$ corresponds to
$B\simeq 6$ Tesla.
In the following numerical calculations,
the hopping parameters are chosen to be $t_x=t_{y}=t_{z}=t$, and
$M_1=t_x+t_y+0.6t_z$.
The last term is the on-site random potential accounting for
disorder scattering, where $w_{n,m,l}$ is uniformly distributed
in the range $w_{n,m,l}\in \lbrack -W/2,W/2]$,
with $W$ as the disorder strength~\cite{Huo1992,Sheng1997}.

In the linear response regime, the charge current in response to an
electric field and a temperature gradient can be written as  ${\bf J}
= {\hat \sigma} {\bf E} + {\hat \alpha} (-\nabla T)$, where ${\hat
\sigma}$ and ${\hat \alpha}$ are the electrical and thermoelectric
conductivity tensors, respectively.
The electrical conductivity $\sigma_{ij}$
at zero temperature can be calculated by using the Kubo formula~\cite{Wang2017}
\begin{eqnarray}
\sigma _{ij}= \frac{ie^{2}\hbar}{S}\sum_{{\epsilon
_{\alpha}}\neq{\epsilon _{\beta}}}\frac{f(\epsilon_\alpha)-f(\epsilon_\beta)}{\epsilon_\alpha-\epsilon_\beta}\frac{\langle
\alpha\mid V_i\mid\beta\rangle\langle\beta\mid V_j\mid\alpha
\rangle}{\epsilon_\alpha-\epsilon_\beta+i\eta}.
\end{eqnarray}
Here, $i,j=x,y$, and $\epsilon_\alpha$ and $\epsilon_\beta$ are the eigenenergies
corresponding to the eigenstates $\vert\alpha\rangle$ and
$\vert\beta\rangle$ of the system, respectively,
which can be obtained through exact diagonalization of
the Hamiltonian Eq.\ (2).
$S$ is the cross-section area in the $x$-$y$ plane.
$f(\epsilon_\alpha)$ and $f(\epsilon_\beta)$ are the Fermi-Dirac
distribution functions, defined as $f(x) = 1/[e^{(x-E_F)/k_B T}+1]$.
$V_{i}$ and $V_{j}$ are the velocity operators, and $\eta$ is
the positive infinitesimal. The Hall conductivity $\sigma_{xy}$,
as the summation of contributions from all layers,
has a dimension of $e^2/h$.

\begin{figure}[tbh]
\includegraphics[width=3.1in]{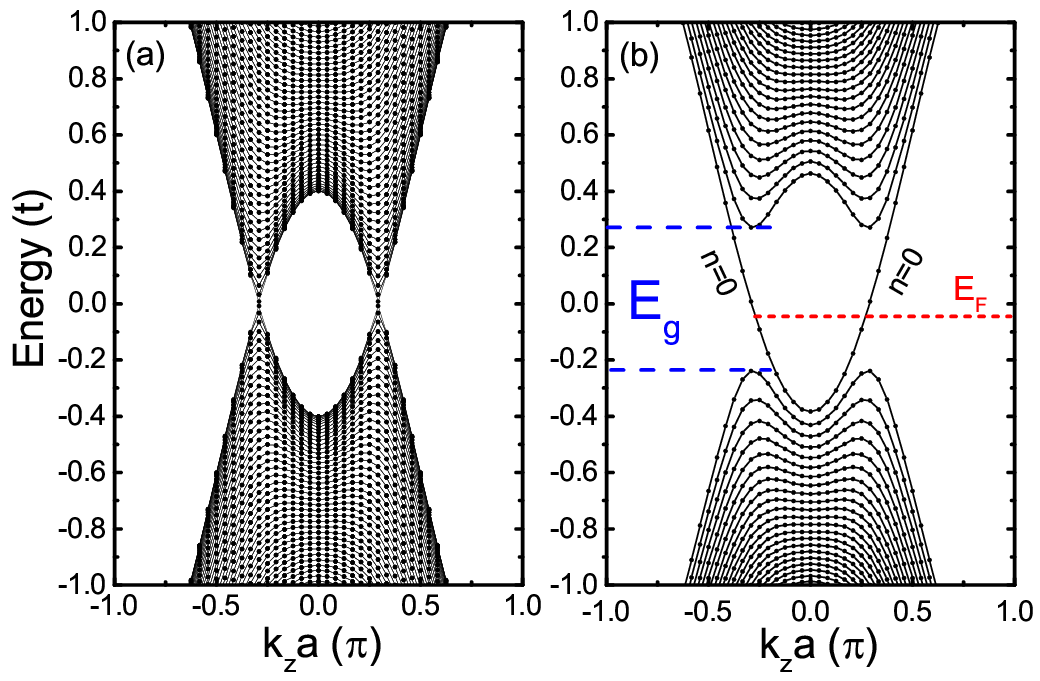}
\caption{ (color online). The energy dispersion as a function of
the wave vector $k_z$ with a periodic
boundary condition for a 3D Weyl semimetal.
(a) $\phi=0$, (b) $\phi=1/192$.
The system size in the $x$ direction is taken to be $n_x=192$.
The zeroth Landau levels (LLs) are labeled by $n=0$.
The gap between $n=-1$ and $n=1$ LLs is labeled
as $E_g$. The Weyl point is set as zero of the energy.
} \label{fig.1}
\end{figure}

\begin{figure}[tbh]
\includegraphics[width=3.1in]{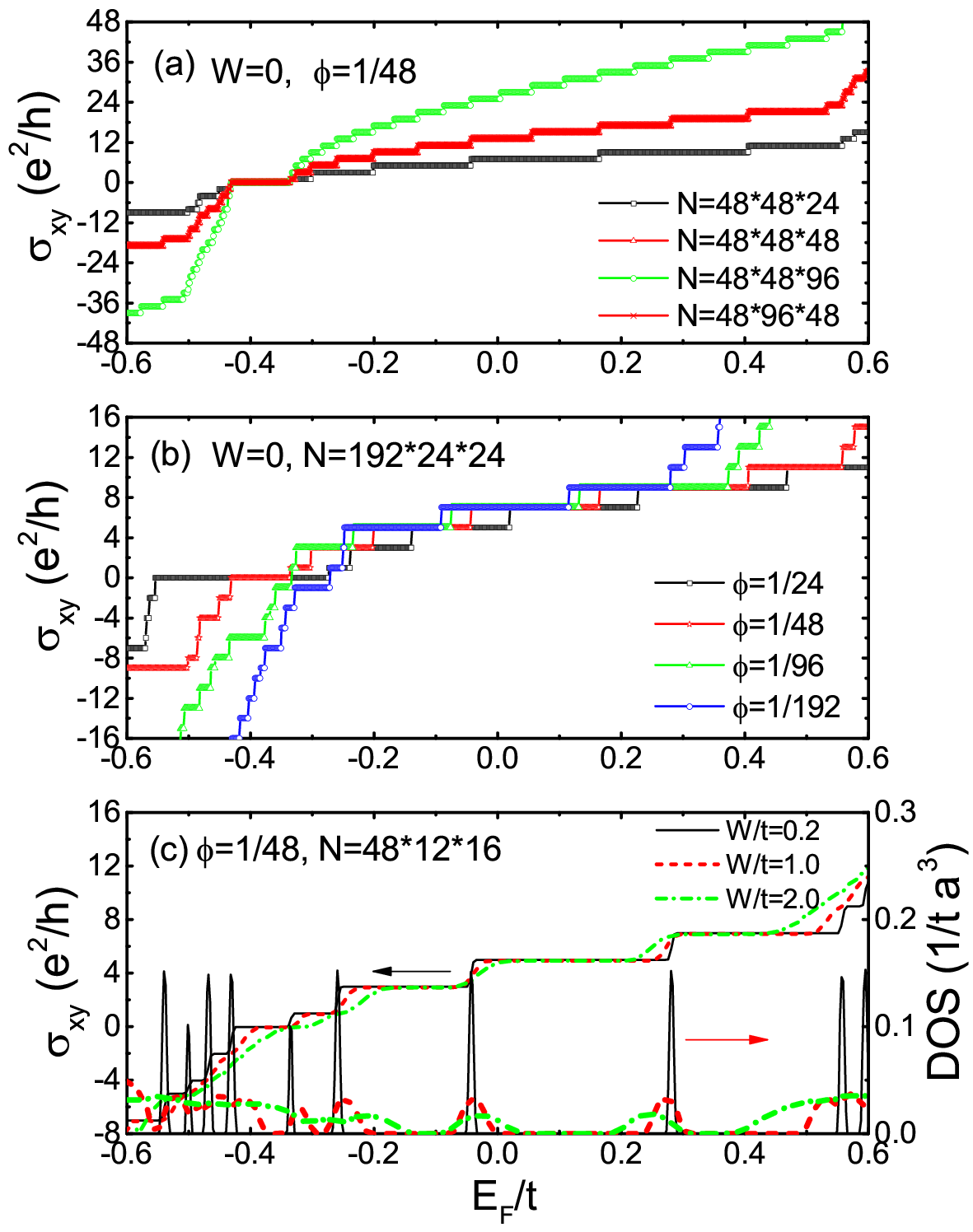}
\caption{ (color online). Calculated Hall conductivities $\sigma_{xy}$
near the band center in Weyl semimetal.
(a) For the system size dependence of $\sigma_{xy}$.
Here, the magnetic flux is chosen as $\phi=1/48$.
(b) For the magnetic field dependence of $\sigma_{xy}$.
Here, the system size is fixed at $N=192\times 24\times 24$,
the magnetic flux is chosen as $\phi=1/24$, $1/48$,
$1/96$ and $1/192$, respectively.
In (a) and (b), the disorder strength is set to $W=0$.
(c) For the disorder effect on $\sigma_{xy}$ and DOS.
The disorder strength is chosen to be $W/t=0.2$, $1.0$ and $2.0$, respectively.
The system size is taken to be $N=48\times 12\times 16$, and
the magnetic flux is chosen as $\phi=1/48$.} \label{fig.2}
\end{figure}

\begin{figure}[tbh]
\includegraphics[width=3.2in]{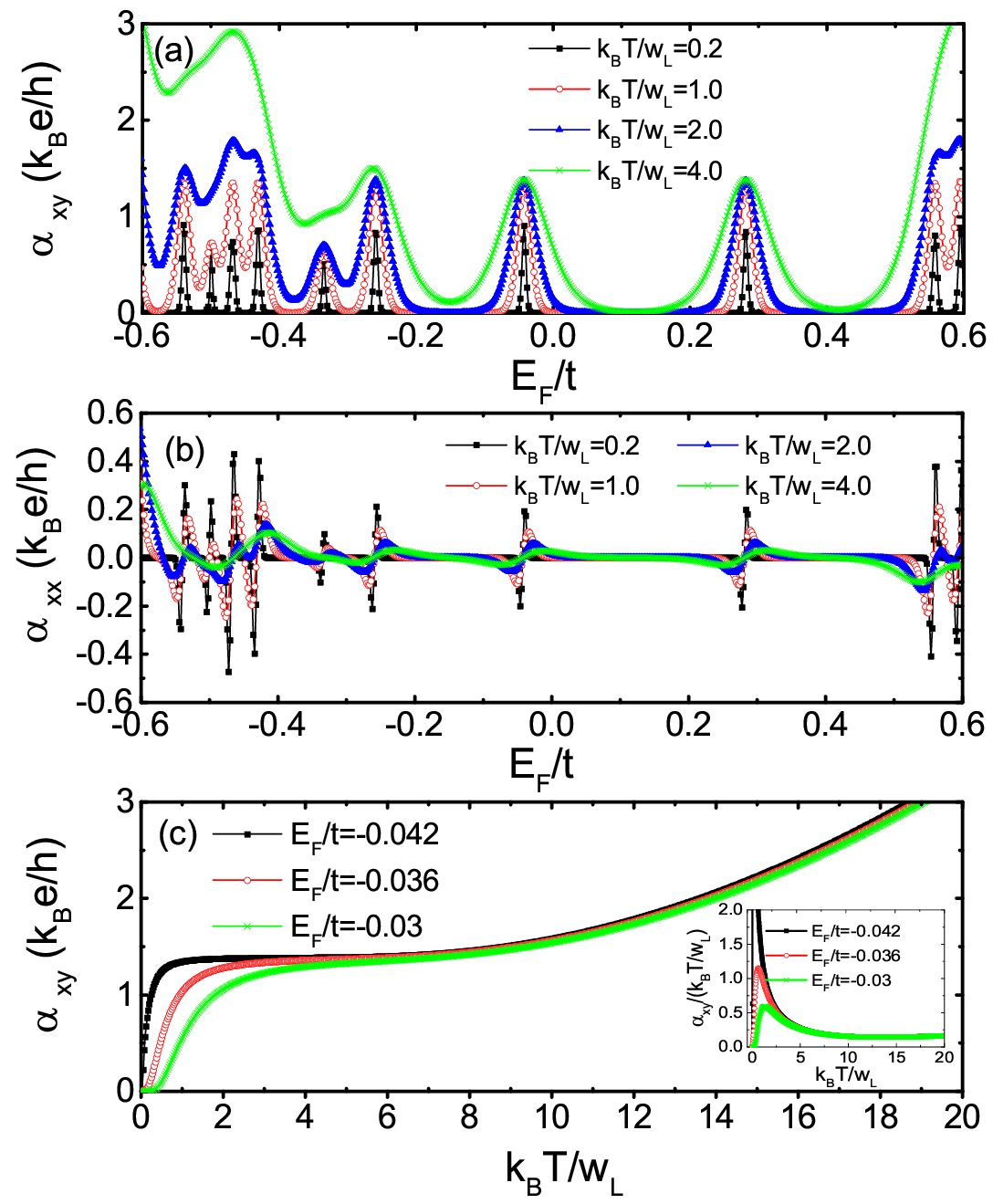}
\caption{ (color online). Thermoelectric conductivities at finite temperatures.
(a) $\alpha_{xy}$, (b) $\alpha_{xx}$
as functions of the Fermi energy at different temperatures.
(c) shows the temperature dependence of $\alpha_{xy}$ for certain fixed Fermi energies.
The inset shows $\alpha_{xy}/(k_BT/W_L)$ as a function of the temperature.
Here, $W_L$ is chosen as $W_L/t=0.005$,
the system size is chosen as $N=48\times 12\times 16$,
the disorder strength is set to $W/t=0.2$, and
the magnetic flux is chosen as $\phi=1/48$.
} \label{fig.3}
\end{figure}

\section{Results and Discussion}

\subsection{ Quantized Hall conductivity}

We first present the energy dispersion with a periodic boundary condition
in the 3D Weyl semimetal.
As shown in Fig.\ \ref{fig.1}(a),
in the absence of a magnetic field,
the conduction and valence bands touch each other
at a pair of Weyl points $k_z=\pm k_0$.
Around these Weyl points the energy dispersion is linear.
When a perpendicular magnetic field is applied, as shown in Fig.\ \ref{fig.1}(b),
the energy spectrum is quantized into the continuum
LLs, except for the chiral zeroth LLs (the curve labeled by $n=0$),
which are separating from the continuum spectrum.
The $n=0$ LLs are apparently discretized, because $k_{z}$ is
quantized owing to the finite thickness $n_{z}a$ in the $z$ direction.
The energy spacing between two neighboring
$n = 0$ LLs is approximately given by
$\Delta E = v_F^z\hbar (2\pi/n_za) = 2\pi t\sin(k_0a) /n_z$ inversely proportional to the thickness $n_za$,
which will determine the width of the quantized Hall plateaus.

We now present the  Hall effect of Weyl semimetals
at zero temperature in the presence of a perpendicular magnetic field.
In Fig.\ \ref{fig.2}(a), the Hall conductivities $\sigma_{xy}$
are plotted as functions of the electron Fermi energy $E_F$
for a clean sample $W=0$ with different system sizes
at the same magnetic flux  $\phi=1/48$.
$\sigma_{xy}$ shows a series of relatively wide quantized
Hall plateaus in units of $e^{2}/h$,
as long as $E_F$ is in the gap between $n=-1$ and $n=1$ LLs of
width about $E_g = 2\sqrt{2}\hbar \omega_c = 2\sqrt{2}\hbar v_F/l_B$,
with $l_B = \sqrt{\hbar/{eB}}$ as the magnetic length~\cite{deng2019}.
For the hopping parameters chosen, the Fermi velocities in the $x$-$y$ plane are
isotropic, so we denote $v_{F}^{x}=v_{F}^{y}=v_{F}$.
The gap $E_{g}$ has been indicated in Fig.\ 1(b).
For $\phi = 1/48$, $E_g = 2\sqrt{2}\hbar v_F/\sqrt{\hbar/{eB}} = 4\sqrt {\pi\phi}t \simeq 1.0 t$.
Outside the gap $E_{g}$, we  see  much narrower  plateaus,
originating from energetically overlapped multiple
LL subbands, which are unstable when we turn on the random disorder.
The quantized Hall conductivity displays a pronounced
electron-hole asymmetry  due to
the asymmetry of the band structure.
When the system size in the $z$ direction is increased
from $n_z=24$ to $n_z=96$,
the Hall conductivity remains to show
the quantized plateaus in units of $e^{2}/h$, but the width
of the wide Hall plateaus around the band center,
determined by $\Delta E=2\pi t\sin(k_0a) /n_z$, decreases from about $0.2t$ to
$0.05t$, proportionally to $1/n_{z}$.
In the limit of infinite $n_{z}$,
the Hall conductivity will lose quantization,
as the energy spectrum of the $n=0$ LLs also becomes continuous.
However, all the results of $\sigma_{xy}$ remain unchanged by changing the system sizes
in the $x$-$y$ plane. For example,
in Fig.\ \ref{fig.2}(a), the calculated Hall conductivities for $n_{y}=48$ and
$n_{y}=96$ collapse into the single red curve.

In Fig.\ \ref{fig.2}(b), we present the Hall conductivities
with different magnetic field strengths for a clean sample $W = 0$.
The system size is fixed at $N=192\times 24\times 24$.
As we can see, more wide quantized Hall plateaus
emerge, as the gap between the $n=-1$ and $n=1$ LLs
increases from $E_{g}=4\sqrt {\pi\phi}t\simeq 0.5t$ to $E_{g}=1.4t$ with
increasing magnetic flux from $\phi=1/192$ to $1/24$.
In Fig.\ \ref{fig.2}(c),
we show the effect of random disorder on the  QHE
and electron density of states (DOS) for fixed system size and magnetic flux.
It is found that
the wide plateaus with $\nu = 3, 5, 7$
around the band center are most robust against disorder scattering.
The disorder affects the plateaus through inducing the LL broadening $\Gamma$~\cite{Jonson1984,ShiPRL2009}.
In the absence of disorder, the DOS is singular for these $n=0$ LLs.
When the disorder is introduced, the DOS is broadened showing a series of peaks around each LL.
The magnitude of $\Gamma$ can be estimated from the width of the broadened DOS.
For example, $\Gamma\simeq 0.1t$ for $W=2t$.
A finite temperature $k_BT$ plays a similar role to $\Gamma$.
The quantized Hall plateaus will remain stable until the LL broadening $\Gamma$ or temperature
$k_{B}T$ becomes comparable to the energy spacing $\Delta E$ between neighboring $n=0$ LLs.
For both ZrTe$_5$ and NbP, the Fermi velocity is about $v_{F}^{z}\sim 5\times 10^5m/s$~\cite{ZrTe5PNAS2017,NbPNatPhys2015}.
We can estimate $\Delta E\simeq 0.02eV$, or equivalently, $230K$ in temperature for a
$n_{z}a=100nm$ thick system.

\begin{figure}[tbh]
\includegraphics[width=2.8in]{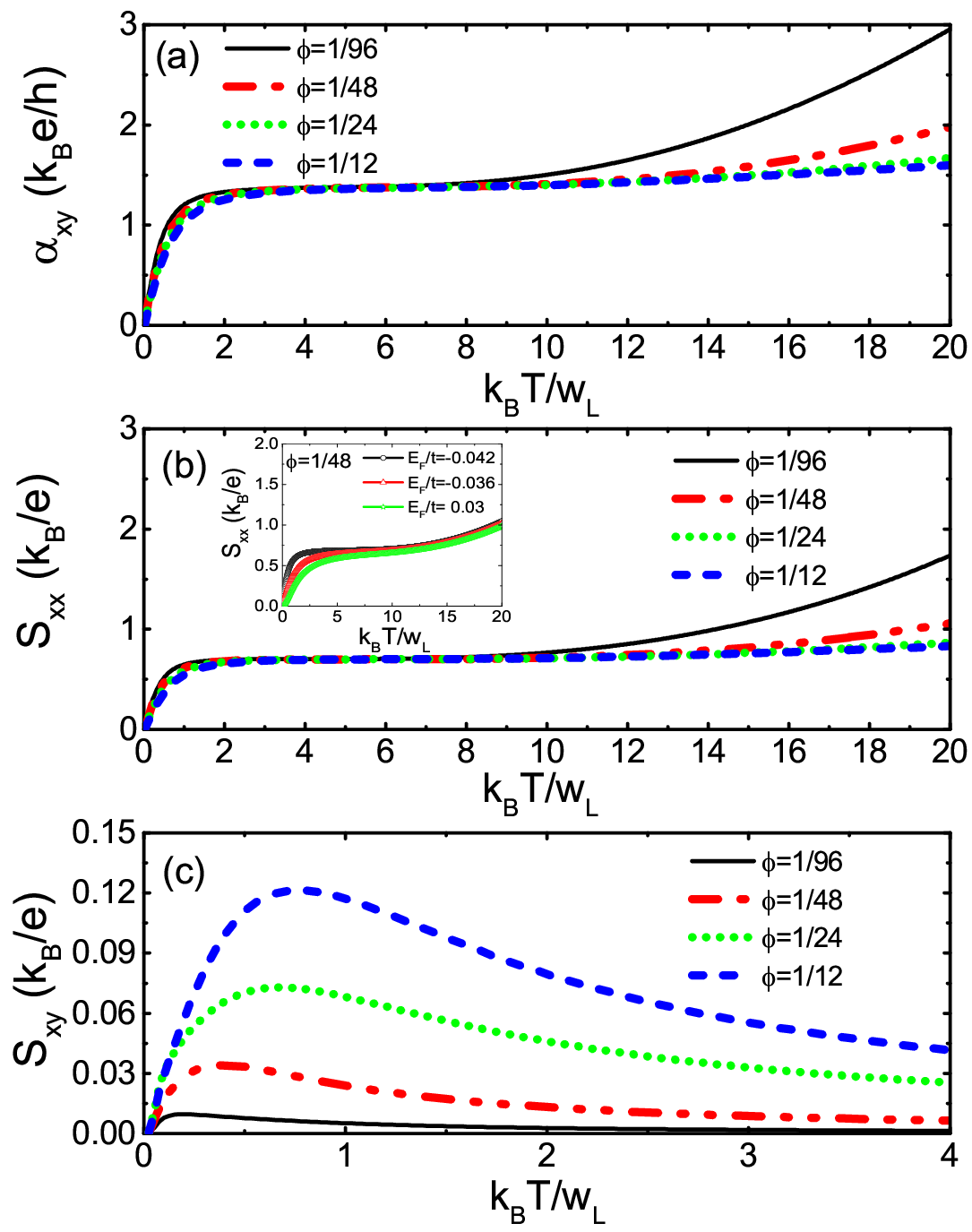}
\caption{ (color online). The temperature dependence of
(a) $\alpha_{xy}$, (b) $S_{xx}$, and (c) $S_{xy}$ at central LL
for different magnetic flux $\phi=1/96$, $1/48$,
$1/24$, and $1/12$, respectively.
The inset of (b) shows the temperature dependence of $S_{xx}$
for certain fixed Fermi energies.
The system size is chosen as $N=192\times 4\times 8$,
the disorder strength is set to $W/t=0.2$.
} \label{fig.4}
\end{figure}

\subsection{ Thermoelectric transport}

Now we turn to the disorder effect on the thermoelectric
transport coefficients in the presence of the strong magnetic field~\cite{supplement}.
In Fig.\ \ref{fig.3},
we first plot the calculated thermoelectric conductivities
at some finite temperatures.
Here, the temperature dependence is
shown as a function of the ratio between $k_BT$ and $W_L$,
where $W_L$  represents the full-width at the half-maximum
of the longitudinal conductivity $\sigma_{xx}$ peaks around zero energy.
As shown in Figs.\ \ref{fig.3}(a)-(b),
the transverse thermoelectric conductivity $\alpha_{xy}$ displays
a series of peaks, while the longitudinal thermoelectric conductivity
${\alpha_{xx}}$ undergoes a sign reversal and approaches zero
at the center of each LL.
In Fig.\ \ref{fig.3}(c), we show $\alpha_{xy}$ as a function
of the temperature for different Fermi energies.
At low-temperature region,
$k_{B}T \ll W_L$,  $\alpha_{xy}$ increases quickly.
When $k_{B}T$ becomes comparable to or greater than $W_L$,
$\alpha_{xy}$ for all Fermi energies reaches
a constant value $1.38 k_B e/h$, which matches exactly
the universal value $g(\ln 2)k_Be/h$ predicted for 2D QHE systems~\cite{Jonson1984,UnivJPC1984},
with degeneracy $g=2$. This  robust flat plateau feature demonstrates the LL quantization (only states
within the  degenerating LLs contribute to the $\alpha_{xy}$),
which can be probed in experimental  measurements at low temperatures.
However, in the high-temperature region, when $k_{B}T \gg W_L$,
the value of $\alpha_{xy}$ continues to rise with increasing
temperature without saturation.
This non-saturating property at higher temperatures
survives even when the quantized Hall plateaus disappear with $n_z \rightarrow \infty$~\cite{Fu2019,Skinner2018}.
Furthermore, we show quantitative behavior of the ratio of the $\alpha_{xy}$ versus
the normalized temperature $k_BT/W_L$ in the inset of Figs.\ \ref{fig.3}(c).
As we can see, $\alpha_{xy}/(k_BT/W_L)$ curves collapse into
a constant plateau at high temperature.

We further demonstrate some interesting features of thermoelectric coefficients
for different magnetic field strengths.
As seen from Fig.\ \ref{fig.4}(a),
we first plot $\alpha_{xy}$ as a function of the normalized temperature $k_BT/W_L$
with increasing magnetic field strength from $\phi=1/96$ to $\phi=1/12$.
At relatively low temperature regions, all the curves of $\alpha_{xy}$
approach a constant value about $1.38 k_B e/h$.
With increasing temperature,
$\alpha_{xy}$ for different magnetic field strengths
all increase gradually. Interestingly, the  weaker the magnetic field strength is,
the faster $\alpha_{xy}$ grows.
In Figs.\ \ref{fig.4}(b) and (c),
we show the temperature dependence of the thermopower $S_{xx}$
and Nernst signal $S_{xy}$~\cite{supplement},
which can be directly measured in experiments~\cite{Ong2017}.
At relatively high-temperature region, we also observe the values of
$S_{xx}$ increase quickly with the decrease of the magnetic field strength,
and the values are inversely proportional to $B$.
The inset of Fig.\ \ref{fig.4}(b) shows the temperature dependence of $S_{xx}$
for some different Fermi energies.
With increasing temperature, the peak values from Fermi energies
continue to grow gradually
with temperature without saturation.
We suggest that these striking features can be attributed to
the thermal excitations between different $n=0$ LLs.
In Fig.\ \ref{fig.4}(c),
$S_{xy}$ assumes the Arrhenius form $(1/T)e^{-E_F/k_{B}T}$
with the increase of temperature.
The peak values are also proportional to the magnetic field strength,
i.e., $S_{xy}\propto B$.
When the magnetic field strength increases from $\phi=1/96$ to $\phi=1/12$,
the peak value of $S_{xy}$ reaches  $0.12$ $k_B/e$ ($10.35$ $\mu V/K$),
which is in agreement with the minimum measured value $\sim8$ $\mu V/K$~\cite{Caglieris2018}.
More interestingly,
in these curves $S_{xy}$ shows a broadened maximum around $k_BT=0.5W_L$
for strong magnetic fields $B$,  which shifts to a lower temperature
by decreasing $B$.  This similar maximum has also been observed
in the experiments for  the compound TaP and NbP~\cite{Caglieris2018,Watzman2018}.

\section{Summary}

In summary, we numerically investigate the 3D QHE and thermoelectric
transport properties of Weyl semimetals in the presence of disorder.
When a perpendicular magnetic field is applied,
we observe well-formed Hall plateaus in units of $e^{2}/h$ for an intermediate number of layers
and a finite range of Fermi energy near the Weyl points.
We demonstrate how the system size,  magnetic field strength
and disorder influence the quantized Hall plateaus.
Both $\alpha_{xy}$ and $S_{xx}$  exhibit
non-saturating characteristic features with increasing temperature,
which are robust in the thermodynamic limit.
However, for lower temperatures,
$\alpha_{xy}$  for all Fermi energies reaches a constant  plateau value 1.38$k_Be/h$,
signaling the LL quantization.
Our work provides a clearer understanding of the topological
3D QHE and magnetothermoelectric transport in Weyl semimetals.

\acknowledgments
We acknowledge helpful discussions with Ming-Xun Deng.
This work was supported by the National Natural Science Foundation of China
under Grants No. 11574155 (R.M.) and No. 11974168 (L.S.).
Work done at CSUN was supported by National Science Foundation Partnerships
for Research and Education in Materials (PREM), and Grant No. DMR-1828019 (D.N.S.).

\vskip 5mm
\centerline {\bf APPENDIX}
\vskip 3mm

Following the Eq.(2) in the main text,
we exactly diagonalize the model Hamiltonian
in the presence of disorder~\cite{Sheng1997},
and obtain the transport coefficients by using the energy spectra and wave functions.
In practice, we can first calculate the electrical conductivities $\sigma_{ij}$
at zero temperature, and then use the relation~\cite{Jonson1984,UnivJPC1984}
\begin{eqnarray}
\sigma_{ij}(E_F, T) &=& \int d\epsilon \,\sigma_{ij}(\epsilon)
\left ( - {\partial f(\epsilon) \over \partial \epsilon } \right), \\
\alpha_{ij}(E_F, T) &=& {-1\over eT} \int d\epsilon\,
\sigma_{ij}(\epsilon) (\epsilon-E_F) \left ( - {\partial f(\epsilon)
\over \partial \epsilon } \right), \label{eq:conductance-finiteT}
\end{eqnarray}
to obtain the electrical and thermoelectric conductivities at finite
temperatures. The thermopower and Nernst signal can be
calculated subsequently from~\cite{Ong2017,footnote1}
\begin{eqnarray}
-S_{xx} &=&  { E_x /\mid \nabla T \mid} = -({\rho_{xx}\alpha_{xx}-\rho_{xy}\alpha_{xy}}),\\
S_{xy} &=& { E_y / \mid \nabla T\mid} =({\rho_{xx}\alpha_{xy}+\rho_{xy}\alpha_{xx}}),
\label{eq:thermoelectric}
\end{eqnarray}
where $\rho_{ij}$ is the resistivity tensor.


\begin{references}
\bibitem{footnote1} Different literatures may have a sign difference due to different conventions.
\end{references}
\end{document}